\documentclass[twocolumn,prl,showpacs,superscriptaddress]{revtex4}
\usepackage{graphicx}
\usepackage{dcolumn}
\usepackage{bm}
\begin{document}

\title{Long-Range Coherence in a Mesoscopic Metal near a Superconducting 
Interface}
\author{H. Courtois}
\email{herve.courtois@polycnrs-gre.fr}
\author{Ph. Gandit}
\affiliation{Centre de Recherches sur les Tr\`es Basses 
Temp\'eratures-C.N.R.S. associated to Universit\'e Joseph Fourier, 25 
Ave. des Martyrs, 38042 Grenoble, France}
\author{D. Mailly}
\affiliation{Laboratoire de Microstructures et de Micro-\'electronique, 196 Av. H. 
Ravera, 92220 Bagneux, France}
\author{B. Pannetier}
\affiliation{Centre de Recherches sur les Tr\`es Basses 
Temp\'eratures-C.N.R.S. associated to Universit\'e Joseph Fourier, 25 
Ave. des Martyrs, 38042 Grenoble, France}
\date{1995, July}
\begin{abstract}
We identify the different contributions to quantum interference in a 
mesoscopic metallic loop in contact with two superconducting 
electrodes. At low temperature, a flux-modulated Josephson coupling 
is observed with strong damping over the thermal length $L_{T}$. At higher 
temperature, the magnetoresistance exhibits large  $h/2e$-periodic 
oscillations with $1/T$ power law decay. This flux-sensitive 
contribution arises from coherence of low-energy quasiparticles 
states over the phase-breaking length $L_{\varphi}$. Mesoscopic fluctuations 
contribute as a small $h/e$ oscillation, resolved only in the purely 
normal state.
\end{abstract}
\maketitle

In a disordered metal at low temperature, electronic coherence
persists over the phase-breaking length $L_{\varphi}$ \cite{AAS}. 
Weak localization, which consists in electron coherent backscattering
along a closed diffusion path, induces corrections of the conductance
of order the quantum of conductance $e^2/h$.  The sensitivity of this
process to an Aharonov-Bohm flux leads to $\phi_{0}=h/2e$ periodic
oscillations of the resistance of a mesoscopic loop
\cite{Sharvins,Pannetier}.  Hybrid systems made of Normal (N) and
Superconducting (S) materials are the scene for new physics, due to
the Andreev reflection and the proximity effect.  At low temperature
($k_{B}T \ll \Delta$), incident electrons have an energy much smaller
than the gap $\Delta$ of S and are Andreev-reflected at the N-S
interface into a coherent hole. Spivak and Kmelnitskii
investigated the effect of Andreev reflection on weak localization in
a S-N-S geometry \cite{Spivak}.  The N metal conductance was predicted
to be sensitive to the phase difference between the two
superconductors with a period of $\pi$, leading to a $h/4e$
flux-periodicity in a loop.  Petrashov et al.  \cite{Petrashov} and de
Vegvar et al.  \cite{Vegvar} measured phase-sensitive transport in
mesoscopic N-S metallic systems.  The interpretation of Ref. 
\cite{Petrashov} results in terms of weak localization is not
consistent with the large amplitude of the effect \cite{VegvarComm}. 
In fact, the proximity effect in such mesoscopic systems can lead to a
zero-resistance state with a well-defined Josephson current
\cite{Courtois} if N-S interfaces have high transparency.  In a
two-dimensional electron gas, Dimoulas et al.  also observed, beyond
the Josephson coupling, large effects of quasiparticle interference on
the resistance \cite{Dimoulas}.  Recently, there has been considerable
interest in coherent transport through mesoscopic N-S tunnel junctions
\cite{Kastalsky,Beenakker}.  Confinement of electrons and holes by
disorder in N induces coherent multiple Andreev reflections, which
enhance the low-temperature subgap conductance \cite{Hekking}.  This
is exemplified by the flux-modulation of the subgap current in the
case of a fork-shaped S electrode \cite{Pothier}.  Volkov showed
that this behaviour may be explained by the appearance, despite the
barrier, of a small pair-amplitude in N \cite{Volkov}.  This suggests
that the proximity effect could explain most of the surprising data on
resistive transport in mesoscopic N-S devices, even if classical
estimates fail to agree with experimental results.

At present, a clear identification of the different contributions to
coherent transport in N-S systems is missing.  In this Letter, we
describe new experimental results revealing unambiguously the nature
of the different contributions to the phase-sensitive current in a
mesoscopic N metal with S electrodes.  Our experimental situation is
greatly simplified compared to others, since we consider only the
conductivity of the normal metal, the role of the interface being
restricted to providing Andreev reflection.  We used a ring geometry
similar to that of Ref.  \cite{Petrashov}, in order to control the
phase by an external magnetic field, and made a study as a function of
temperature.  We show under which conditions the Josephson coupling,
weak localization and other contributions prevail.  One important
result is in the intermediate temperature regime, when the thermal
length $L_{T}$ is much smaller than $L_{\varphi}$ and the sample size. 
We find a phase-sensitive contribution with an amplitude described by
a $1/T$ power law.  This contribution, much larger than the weak
localization contribution, results from the persistence of
electron-hole coherence far away from the N-S interface.

\begin{figure}
\includegraphics[width=8.5 cm]{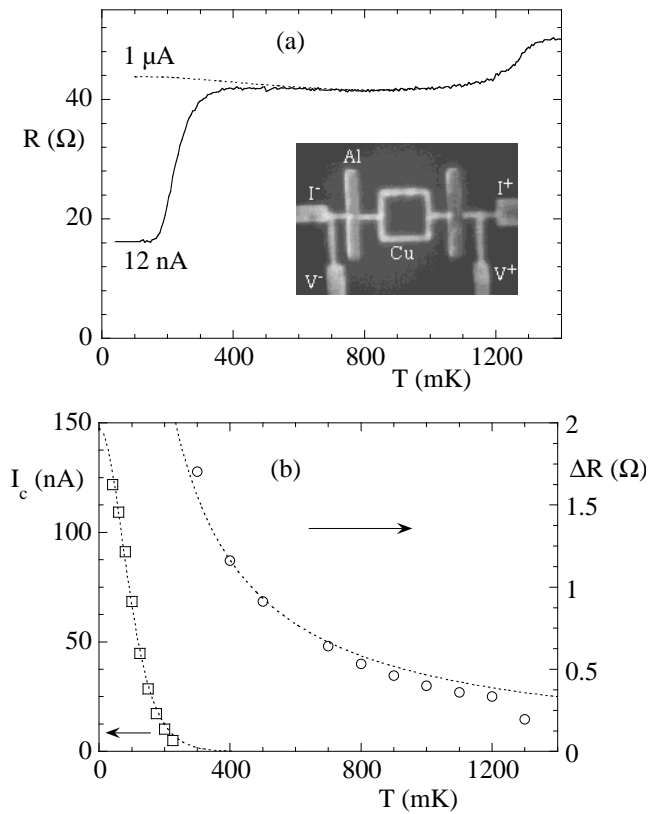}
\caption{(a) Temperature dependence of the resistance with a 
measurement current $I_{mes}$ = 12 nA and 1 $\mu$A. Inset : Micrograph of a 
typical sample made of a Cu square loop with 4-wire measurement 
contacts, in contact with two Al islands (vertical). Diameter of the 
loop is 500 nm, width 50 nm, thickness 25 nm. Centre-to-centre 
distance between the 150 nm wide Al islands is 1 $\mu$m. The length L of 
the N part of the S-N-S junction is 1.35 $\mu$m. (b) Left scale : 
Temperature dependence of the critical current derived from Fig. 2 
data with a 25 $\Omega$ differential resistance criteria. Dashed line is a 
guide to the eye. Right scale : Temperature dependence of the 
amplitude of the magnetoresistance oscillations, $I_{mes}$ = 60 nA. Dashed 
line is a $1/T$ fit.}
\end{figure}

The inset of Fig.  1(a) shows a micrograph of a typical sample, made
of a square Cu loop in contact with two Al electrodes.  The Cu loop
and Al electrodes are patterned by conventional lift-off e-beam
lithography in two successive steps with repositioning accuracy better
than 100 nm.  In-situ cleaning of the Cu surface by 500 eV Ar$^+$ ions
prior to Al deposition ensures us of a transparent and reproducible
interface.  We performed transport measurements in a Mumetal-shielded
dilution refrigerator down to 20 mK. Miniature low temperature
high-frequency filters were integrated in the sample holder
\cite{RSICourtois}.  We focus here on the experimental results of one
sample representative of others.  The 51 $\Omega$ normal-state
resistance gives an elastic mean free path $l_{p}$ of 16 nm and a
diffusion constant $D$ of 81 cm$^2$/s.  The amplitude of the $h/e$
oscillation in the normal state (see below) provides $L_{\varphi}$ =
1.9 $\mu$m, so that the whole structure is coherent.  The much smaller
decay length of the pair amplitude in N is $L_{T}=\sqrt{\hbar D/2\pi
k_{B}T}=99 nm/\sqrt{T(K)}$.  Fig.  1 summarizes our results.  Below
the superconducting transition of Al at $T_{c} \simeq$ 1.4 K, the
resistance of the sample decreases (Fig.  1(a)) by an amount
corresponding to the coverage ratio of the Cu wire by Al islands
($\simeq 20 \%$).  This behaviour has already been met in previous
experiments [8] and takes place provided the N-S interface resistance
is low.  This give us a lower bound of 7 $\%$ for the interface
transparency $t_{0}$ and an upper bound of 230 nm for the
barrier-equivalent length $L_{t} = l_{p}/t_{0}$ \cite{Zhou}.  At very
low temperature (T $<$ 250 mK), the sample resistance drops to a
constant value of 16 $\Omega$.  The suppression of this drop by a
large bias current suggests a Josephson coupling between the two S
electrodes.  The fraction of residual resistance (31 $\%$) can be
related to the resistance of the normal metal between each voltage
contact and the neighbouring S island (23 $\%$), with an extra
contribution due to the current conversion.  Anticipating the
discussion of the results, we show in Fig.  1(b) the temperature
dependence of the two main contributions to transport.  The Josephson
current vanishes rapidly above 250 mK, revealing the exponential decay
over $L_{T}$.  The amplitude of the observed $h/2e$ magnetoresistance
oscillations is plotted on the same graph.  It represents more than 1
$\%$ of the loop resistance and can be reasonably fitted by a $1/T$
power law.  Let us now discuss these observations in more detail.

\begin{figure}
\includegraphics[width=8.5 cm]{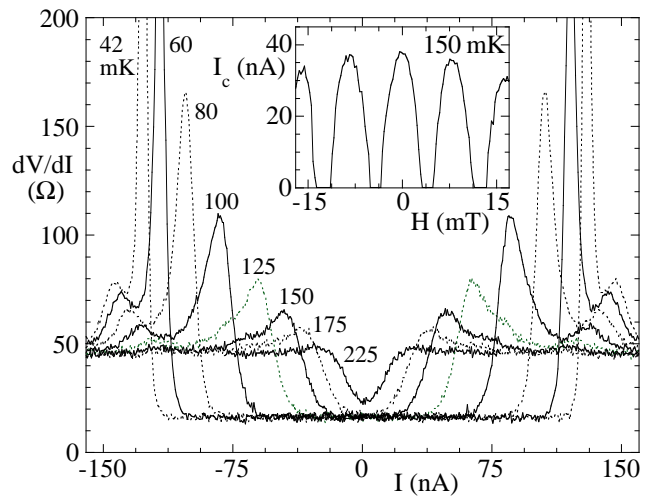}
\caption{Current - differential resistance characteristics at 
temperatures T = 42; 60; 80; 100; 125; 150; 175 and 225 mK, $I_{mes}$ = 3 
nA. Inset: Magnetic field dependence of the critical current at T = 
150 mK of the low temperature, low current R = 16 $\Omega$ resistance 
plateau. Differential resistance criteria is 35 $\Omega$ and $I_{mes}$ = 3 nA. 
The 8.25 mT periodicity of the oscillations corresponds to a quantum 
of flux $\phi_{0} = h/2e$ in the 0.25 $\mu$m$^2$ loop area.}
\end{figure}

Figure 2 shows the current - differentialresistance characteristics for
different temperatures between 42 and 225 mK. A sharply peaked feature
indicates the switching of the loop into a resistive state.  An
unexplained additional structure can be seen at higher current.  At
the highest temperatures, a thermal rounding of the characteristics is
visible.  The shape of these curves can be qualitatively accounted for
in a Resistively Shunted Junction (RSJ) model with thermal
fluctuations \cite{Charlat}.  Solving the linearized Ginzburg-Landau equations,
it is straightforward to calculate the pair current $J_{s}$ between the two
$\varphi$-dephased S electrodes.  If the length L of N metal between the two S
electrodes is large compared to $L_{T}$, we find a sinusoidal current-phase
relation \cite{Note1} : 
\begin{equation}
    J_{s}=J_{0}\exp[-L/L_{T}]\cos(\pi \phi/\phi_{0})\sin\varphi,
\end{equation}
where $J_{0}$ is a constant.  The modulation of the maximum pair
current by the magnetic flux $\phi$ with a period $\phi_{0}$ is
reminiscent of a superconducting quantum interference device (SQUID),
although our mesoscopic geometry differs strongly from the classical
design.  Fig.  2 Inset shows the magnetic field dependence of the
critical current at 150 mK. The 8.25 mT periodicity gives a
flux-periodicity in the 0.25 µm$^2$ loop area, in agreement with (1). 
The observation of a Josephson coupling not found in Ref. 
\cite{Petrashov} is attributed to the high transparency of our N-S
interfaces.

\begin{figure}
\includegraphics[width=8.5 cm]{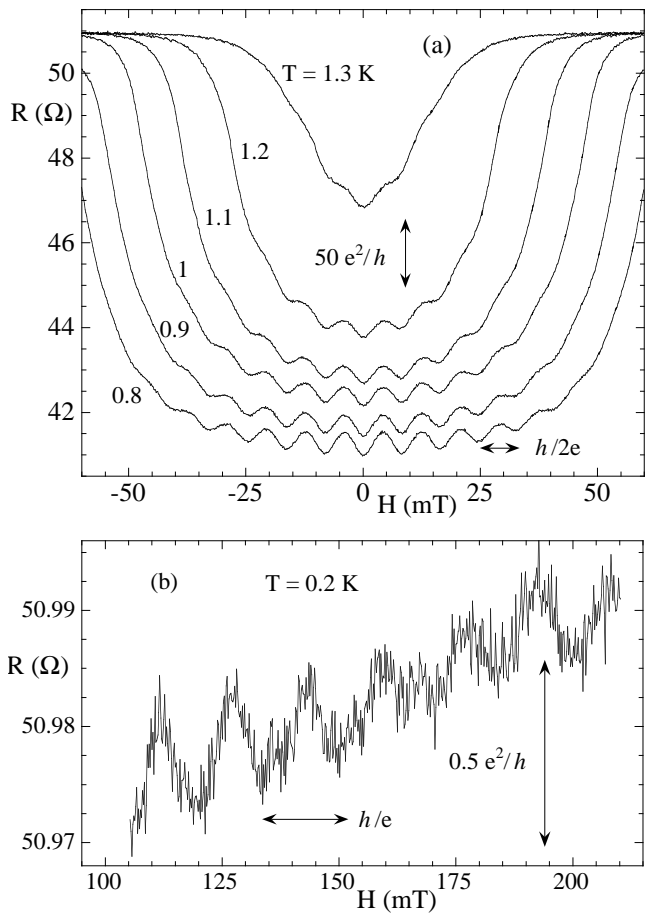}
\caption{(a) : Low-field magnetoresistance for T = 0.8; 0.9; 1; 
1.1; 1.2 and 1.3 K with $I_{mes}$ = 60 nA. T = 0.8 and 0.9 K curves have 
been shifted down by 1 and 0.5 $\Omega$ respectively for clarity. 
Oscillations of periodicity $h/2e$ and amplitude 16.7 $e^2/h$ at T = 0.8 K 
are visible. (b) : High-field, and consequently normal-state, 
magnetoresistance of the same sample at T = 0.2 K, $I_{mes}$ = 600 nA. 
Conductance fluctuations appear, with a main component of periodicity 
$h/e$ and magnitude of order 0.34 $e^2/h$ , which gives $L_{\varphi} 
\simeq 1.9 \mu$m.}
\end{figure}

In the high temperature regime (500 mK $< T < T_{c}$) the
pair current is thermally suppressed.  Calculating the thermal
fluctuations of the pair current in the RSJ model \cite{Bishop}, one expects
exponentially-small magnetoresistance oscillations :
\begin{equation}
    \Delta R/R_{N}=-(1/8)(\hbar J_{s}/e k_{B}T)^2,
\end{equation}
which extrapolates to 10$^{-8}$ at 1 K. In contrast, the
magnetoresistance measured at various temperatures in this regime (see
Fig.  3(a)) shows pronounced oscillations of $h/2e$ periodicity, which
is consistent with Ref.  \cite{Dimoulas}.  With a sample resistance
between the two S electrodes $R_{N}$ = 29 $\Omega$, the relative
amplitude reaches about 1.4 $\%$ at 1 K. No structure of half
periodicity was met when measuring current amplitude was changed [20]. 
The amplitude of the low-field oscillations plotted in Fig.  1(b)
shows a good agreement with a plain $1/T$ fitting law.  The slight
deviation of the data from the $1/T$ fit near $T_{c}$ should be
related to the depletion of the gap in S. This power-law dependence is
a new result, in clear contrast with the exponential damping over
$L_{T}$ of the Josephson current.  The characteristic features of the
observed magnetoresistance oscillations are the following : (i)
precise $h/2e$ periodicity with a resistance minimum at zero-field,
(ii) survival beyond the cut-off $L_{T}$ of the Josephson effect,
(iii) vanishing when the superconductivity of Al is destroyed above
$T_{c}$ or above critical magnetic field, (iv) same effect observed in
samples with only one S island \cite{Charlat}, (v) clear difference
from mesoscopic fluctuations (period $h/e$) and weak localization
($h/2e$) effects, both of amplitude $e^2/h$.  For comparison, Fig. 
3(b) shows the high-field magnetoresistance of the same sample when Al
superconductivity is destroyed.  We observe conductance fluctuations,
with a $h/e$ periodicity and a much smaller amplitude of 0.34 $e^2/h$,
which gives $L_{\varphi} \simeq 1.9 \mu$m.

Our observations appear to be consistent with recent work from Zhou,
Spivak and Zyuzin \cite{Zhou} and previous work by Zaitsev
\cite{Zaitsev} and Volkov et al.  \cite{Volkov,VolkovPrivee}.  Zhou,
Spivak and Zyuzin showed that, beyond the expected strong suppression
of the electric field in N over a length $L_{T}$ from the N-S
interface, corrections to resistive transport survive over distances
up to $L_{\varphi} > L_{T}$.  These corrections follow a power-law
temperature dependence due to the long-range coherence of low-energy
electron-hole pairs.  Indeed, the decay length
$L_{\epsilon}=\sqrt{\hbar D/\epsilon}$ for the pair-amplitude
wave-function $F(\epsilon,x)$ diverges near the Fermi level $\epsilon$
= 0 \cite{Zhou}.  Let us propose a simple picture for this effect.  At
the N-S interface, an incident electron is reflected into a hole of
the same energy $\epsilon$, but with a change in wave-vector $\delta k
= \epsilon/\hbar v_{F}$ due to the branch crossing.  After diffusion
to a distance L from the interface, this induces a phase shift $\delta
\varphi = \delta k v_{F} L^2/D=\epsilon/\epsilon_{c}$ between the
electron and the hole.  This means that at a distance $L > L_{T}$,
electron-hole coherence is restricted to an energy window of width the
Thouless energy $\epsilon_{c}=\hbar D/L^2$ which is small compared to
the width $k_{B}T$ of the thermal distribution.

Resistive measurements in mesoscopic systems are strongly sensitive to
the nature of electrical probes and the location of electron
reservoirs, so that a quantitative description of the resistance
behaviour requires consideration of the particular geometry of the
sample with its contacts.  Nevertheless, the location of the N loop at
a distance significantly larger than $L_{T}$ from the N-S interface
enables us to select the long-range component of the coherent
quasiparticles states.  The magnetic field creates an additional
phase-shift of $2 \pi$ between the electron and the hole at flux
$h/2e$, leading to $h/2e$-periodic modulation of the resistance. 
Here, we do not observe an oscillation amplitude proportional to
$L_{T} \propto 1/\sqrt{T}$ which would be the contribution of the whole N conductor
between reservoirs \cite{Zhou}, but a $1/T$ dependence related to the
local contribution of the loop.  This phase-sensitive contribution
comes from the excess conductivity of those electron-hole pairs which
have an energy close to $\epsilon_{c}$.  Both the temperature dependence and the
amplitude of order 1 $\%$ of the oscillations are indeed consistent
with the ratio $\epsilon_{c}/k_{B}T$.  The similarity of this law with
the expression of the local electric field at $L > L_{T}$ from Zhou et
al.  calls for further clarification.  At very low temperatures ($T <
\epsilon_{c}/k_{B}$), one enters a new regime with the suppression of
the conductance enhancement at low voltage \cite{Stoof}.  This is not observed
here due to the Josephson coupling, but is observed in samples with a
single S island \cite{Charlat}.

In conclusion, we have clearly identified the different components 
of the proximity effect in a mesoscopic metal near a superconducting 
interface. We demonstrated the cross-over between the low-temperature 
Josephson coupling and a phase-sensitive conductance enhancement at 
high temperature. Compared to weak localization, this contribution 
has a larger amplitude and a distinct origin. The observed power-law 
dependence is consistent with the theoretical description of Zhou et 
al. \cite{Zhou} of long-range coherence of low-energy electron-hole pairs. 
Further theoretical progress is needed to have a clear physical 
understanding of the observed phenomena. This new effect could be 
very powerful in investigating phase-breaking effects in magnetic 
nanostructures \cite{PetrFerro} and electron-electron interaction in metals 
\cite{Zhou,Stoof}.

We thank P. Charlat for help in the experiment, J. Chaussy, A. Volkov
and A. Zaikin for stimulating discussions.  We are indebted to F. Zhou
and B. Spivak for clarifying discussions on their paper \cite{Zhou}. 
This work is supported by European Community under contract n¡
CHRX-CT92-0068.

\end{document}